\documentclass[pra,aps,groupedaddress,superscriptaddress,twocolumn,numerical]{revtex4-1}

\usepackage[T1]{fontenc}
\usepackage[utf8]{inputenc}

\usepackage{amssymb}
\usepackage{amsmath}
\usepackage{subfigure}
\usepackage{float}

\usepackage{braket}
\usepackage{nicefrac}               
\usepackage{xfrac}                  
\usepackage{booktabs}				
\usepackage{multirow}				

\usepackage{tikz}
\usetikzlibrary{positioning,decorations.pathreplacing,shapes,calc}
\usepackage{standalone}

\usepackage[detect-family]{siunitx}
\usepackage[colorlinks=true,
linkcolor=blue,
urlcolor=black,
citecolor=blue]{hyperref}

\newcommand{\phd}[1]{\ensuremath{^{}_{#1}}}
\newcommand{\phdt}[1]{\phd{\text{#1}}}
\renewcommand{\subsection}[1]{\textit{#1:}}
\def\poi#1{(#1)}

\newcommand{\region}[1]{\ensuremath{(#1)}}
\def\ie{\textit{i.e.}\,\,}
\def\eg{\textit{e.g.}\,\,}
\newcommand{\menge}[1]{\ensuremath{\mathcal{#1}}}

\DeclareSIUnit\atoms{atoms}
\DeclareSIUnit\Atoms{atoms}

\def\tof{\ensuremath{t\phdt{TOF}}}

\newcommand{\HzPi}[2]{\ensuremath{2\pi\cdot\SI{#1}{#2\hertz}}}
\newcommand{\MHzPi}[1]{\ensuremath{\HzPi{#1}{\mega}}}
\def\coolerIntensity{\ensuremath{C\phdt{3D}}}
\def\coolerDetuning{\ensuremath{\det\phdt{3D}}}
\def\spectroscopy{\ensuremath{\omega\phdt{Spec}}}
\def\coolerdetuning{\ensuremath{\det\phdt{2D}}}
\def\coolerPower{\ensuremath{P\phdt{3D}}}
\def\repumpPower{\ensuremath{P\phdt{Rep}}}
\def\coolerpower{\ensuremath{P\phdt{2D}}}

\begin{document}
	\title{Optimizing Quantum Gas Production by an Evolutionary Algorithm}

	\author{Tobias Lausch}
	\affiliation{Department of Physics and Research Center OPTIMAS, University of Kaiserslautern, Germany}
	
	\author{Michael Hohmann}
	\affiliation{Department of Physics and Research Center OPTIMAS, University of Kaiserslautern, Germany}
	
	\author{Farina Kindermann}
	\affiliation{Department of Physics and Research Center OPTIMAS, University of Kaiserslautern, Germany}
	
	\author{Daniel Mayer}
	\affiliation{Department of Physics and Research Center OPTIMAS, University of Kaiserslautern, Germany}
	\affiliation{Graduate School Materials Science in Mainz, Gottlieb-Daimler-Strasse 47, 67663 Kaiserslautern, Germany}
	
	\author{Felix Schmidt}
	\affiliation{Department of Physics and Research Center OPTIMAS, University of Kaiserslautern, Germany}
	\affiliation{Graduate School Materials Science in Mainz, Gottlieb-Daimler-Strasse 47, 67663 Kaiserslautern, Germany}
	
	\author{Artur Widera}
	\affiliation{Department of Physics and Research Center OPTIMAS, University of Kaiserslautern, Germany}
	\affiliation{Graduate School Materials Science in Mainz, Gottlieb-Daimler-Strasse 47, 67663 Kaiserslautern, Germany}
	
\begin{abstract}
	We report on the application of an evolutionary algorithm (EA) to enhance performance of an ultra-cold quantum gas experiment.
	The production of a $^{87}$Rubidium Bose-Einstein condensate (BEC) can be divided into fundamental cooling steps, specifically magneto optical trapping of cold atoms, loading of atoms to a far detuned crossed dipole trap and finally the process of evaporative cooling. The EA is applied separately for each of these steps with a particular definition for the feedback the so-called fitness. We discuss the principles of an EA and implement an enhancement called differential evolution.
	Analyzing the reasons for the EA to improve \eg, the atomic loading rates and increase the BEC phase-space density, yields an optimal parameter set for the BEC production and enables us to reduce the BEC production time significantly. Furthermore, we focus on how additional information about the experiment and optimization possibilities can be extracted and how the correlations revealed allow for further improvement.	
	Our results illustrate that EAs are powerful optimization tools for complex experiments and exemplify that the application yields useful information on the dependence of these experiments on the optimized parameters.
\end{abstract}

\keywords{Ultracold Quantum Gases -- Bose-Einstein Condensate -- Optimization -- Evolutionary Algorithm -- Differential Evolution}

\maketitle


\section{Introduction}
\label{sec:introduction}

Quantum gases have proven to be versatile model systems to investigate intriguing phenomena of quantum physics, including, for example, superfluidity in ultracold bosonic \cite{Desbuquois2012} or fermionic quantum gases \cite{Zwierlein2005}.

In an experimental realization, quantum gas production requires several distinct cooling steps, typically including different laser cooling stages, a transfer to a conservative trap, and evaporative cooling to the quantum degenerate regime \cite{Ketterle1999}. 
While these individual steps and the underlying physics are well understood, the experimental apparatus usually requires controlling dozens of parameters. 
These are normally constrained depending on the specific experimental scenario, such as maximum currents or laser powers. 
Beyond the large number of parameters, an additional complication arises from the emergence of correlations between experimental parameters which are sometimes unknown and originate from the particular experimental setup. 
A simple and well-known example for such a correlation is the change of atomic transition frequencies in deep optical traps due to the AC Stark shift \cite{Barber2008,Rosenbusch2009}, which correlates the detuning of external resonant lasers to the intensity that defines the trap depth.
In normal operation, therefore, it is unclear, if an apparatus is operating at the optimum of its capabilities. 
Operating at the optimum, however, is advantageous for statistics of the measurements, stable operation over long times and achieving the fastest possible experimental cycle duration.
We show that, in all cases considered, a manually optimized set of parameters can be improved by application of an evolutionary algorithm.

Evolution is the most natural way of optimization in a given scenario, which in terms of biology may be associated with the evolution of genetic material \cite{Stern2013}.
The corresponding digitized version in information technology is called  \emph{evolutionary algorithm} (EA).
In fact, EAs are commonly applied in various fields \cite{Baumert1997,Pearson2001,Tsubouchi2008,Roslund2009} and moreover have created their own field of research in information technology that is closely linked to swarm optimization algorithms, artificial intelligence and self-teaching technologies \cite{Picard2008,Kennedy1995,Fogel1966}.
The application of such an algorithm has proven to result in interesting improvements, \eg the design of antennas \cite{Hornby2011}.
In general, it is an iterative modification and selection process that optimizes a so-called fitness value on a given set of parameters, \eg selecting the combination of parameters in our quantum gas experiment that leads to the fastest sequence to produce a BEC.

We report on the implementation of an evolutionary algorithm which has  been shown to operate in systems with large dimension that exhibit unknown correlations \cite{Price2005}. The application of such algorithms elucidates these correlations and improves, \eg  the performance of quantum gas production.
This technology is applied to the creation of an all optical BEC of $^{87}$Rubidium (Rb), which in our project is combined with individual $^{133}$Caesium (Cs) atoms.
In order to obtain small statistical errors for measurements on a single atom we require many repetitions and therefore the fastest possible creation cycle for the BEC.
In our setup, approximately 70 parameters need to be controlled during a single realization of the experiment. Controlling and optimizing this number of parameters manually is a challenging and demanding task. Since manual adjustment mostly optimizes each parameter separately, it falls short behind the system's technical potential. We find that using the EA on computer-controlled parameters leads to a general improvement of a pre-defined figure of merit and allows reducing the time for quantum gas production from eight to four seconds.
Previous work has applied an EA to the creation of ultra cold quantum gases and increased the mean atom number of an optical molasses \cite{Rohringer2008} with a three-dimensional set of parameters or increased the phase-space density of a four-dimensional problem by improving the radio-frequencies (RF) and timing of RF-cooling ramps \cite{Rohringer2011}. In a related work the scaling with dimensionality of an EA was analyzed and the introduction of a limited lifetime effectively shown to increase robustness against signal noise \cite{Geisel2013}.
There are also other techniques such as machine-learning algorithms applied to the shaping of the evaporation scheme to increase phase-space density of a BEC \cite{Wigley2015}.
Here, we do not only describe our implementation of such an EA but also give examples how to further enhance the bare EA. We keep focus on the information that can be obtained from applying such an algorithm revealing important correlations and knowledge about the experiment.


\section{The Evolutionary Algorithm}
\label{sec:evolution}
\begin{figure}[t]
	\subfigure[]{
		\includegraphics[width=.45\textwidth]{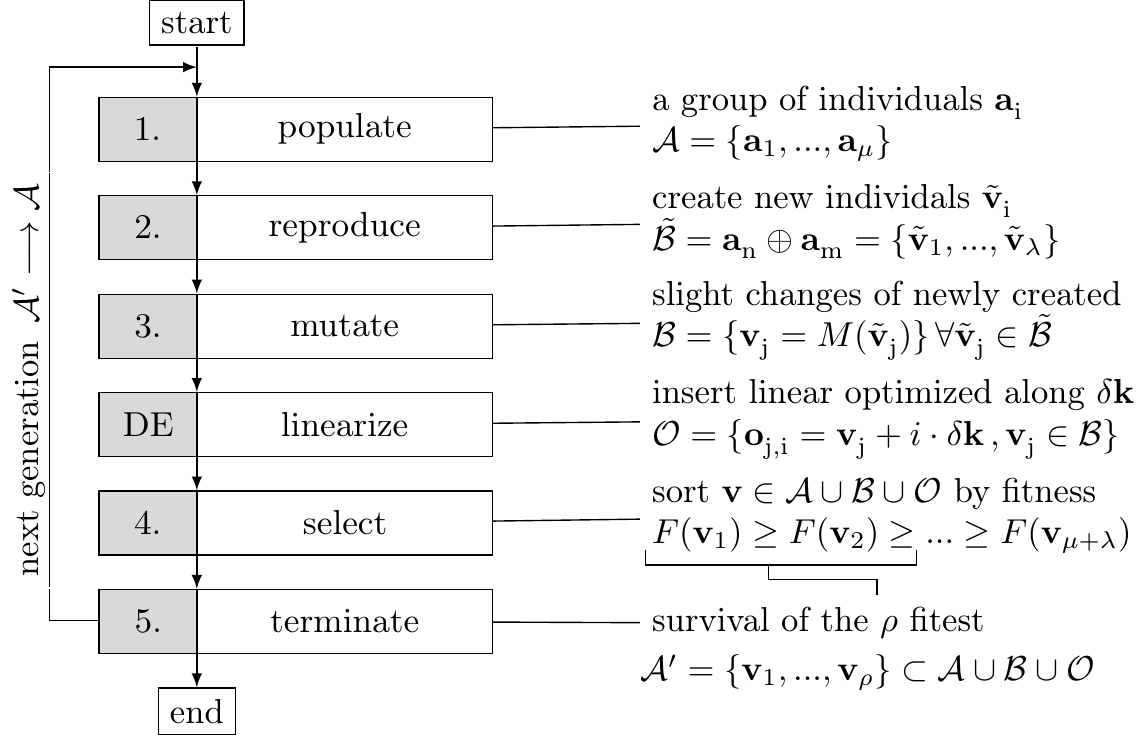}
		\label{fig:evolution_cycle}	
	}
	\subfigure[]{
		\includegraphics[width=.45\textwidth]{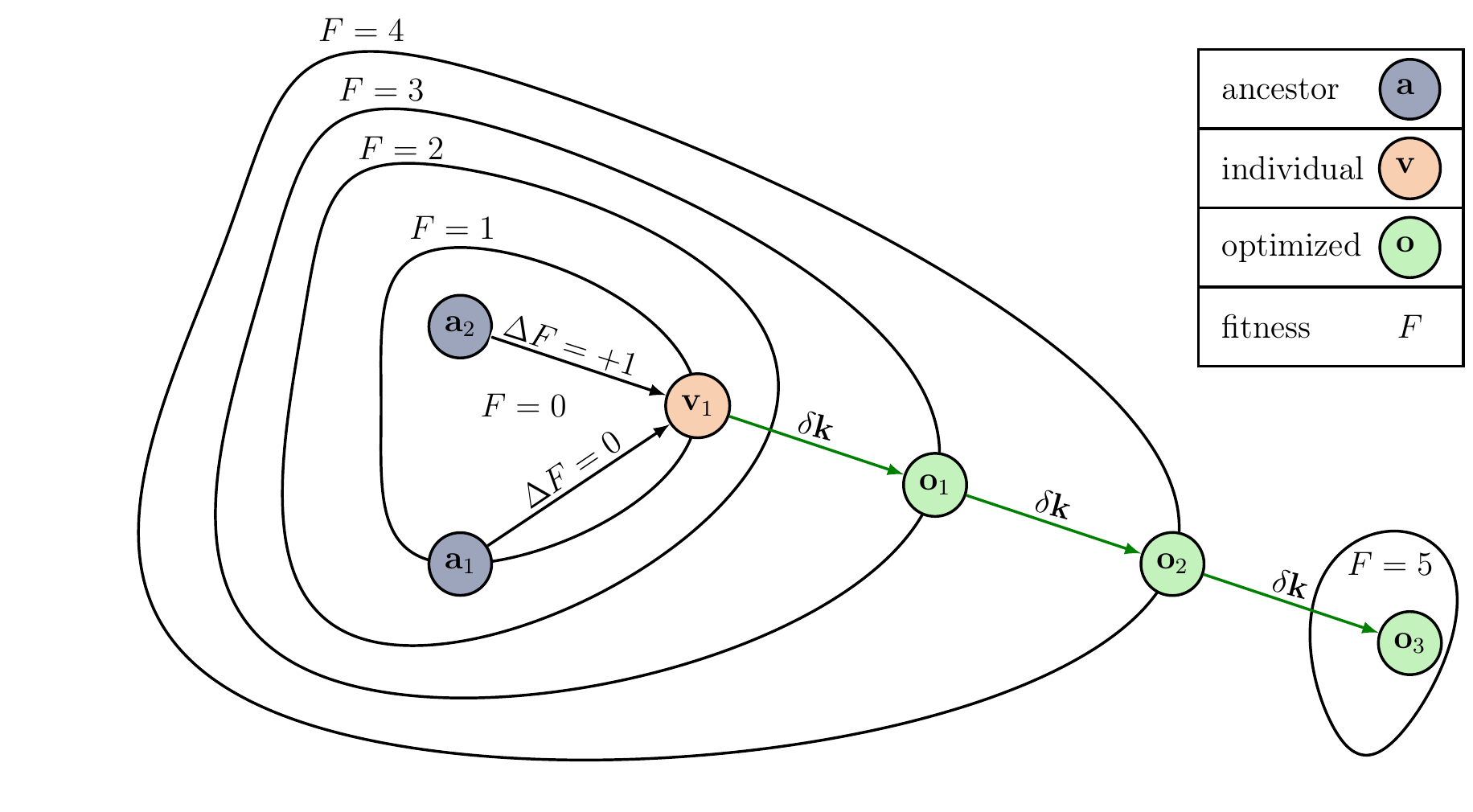}
		\label{fig:steepest_descent}
	}
	\caption{
		EA Scheme: \poi{a} The different processes of an EA (1.-5.) and the integration of differential evolution.
		The algorithm includes reproduction of a population by combining individuals of highest fitness (2.), mutation (3.) of individuals with a certain rate and determination of the fitness. Based on the results of the preceding generation a subsequent steepest descent process adds linearly optimized individuals into the current population.
		Finally the individuals with highest fitness are selected (4.) to become ancestors in the following time step or the evolution is terminated due to having fulfilled a termination condition (5.).
		\poi{b} Example of a steepest descent optimization based on an artificial fitness landscape for different fitness values $F$.
		The improvement of $\Delta F = +1$ from the individual $v$ to its ancestor $a\phd{2}$ creates a weighted optimization vector $\delta \vec{k}$ and multiple optimization candidates $o\phd{1-3}$.
	}
\end{figure}

The steps of the EA implemented  are depicted in Figure  \ref{fig:evolution_cycle}.
In order to quantify the performance of a specific set of parameter values, referred to as an individual $\vec{\nu}\phdt{i} \in \menge{M} $ out of parameter space $\menge{M}$,  it is assigned a single real value as fitness $F: \menge{M} \rightarrow \mathbb{R}$ during the iteration. All individuals of a given time step form a population $\menge{A}$ (step 1).
From the selected individuals of the previous generation, a new population $\menge{B}$ is created in a reproduction step (step 2).
For this second step, we have implemented discrete inheritance $\vec{\nu}\phdt{i} \oplus \vec{\nu}\phdt{j}$, where the partners $\vec{\nu}\phdt{i,j}$ are selected by descending fitness.

Thus the parameters of new individuals match the parameter values of their parents and cannot leave a discontiguous parameter space.
For all applications here the number of parents was chosen to be two, but in general can be adjusted freely. The newly created individuals are subsequently mutated $\vec{c}\phdt{j} = M(\vec{\nu}\phdt{j})$ (step 3), thus their parameter values are randomly changed with a certain probability and within a predefined range.
The probability for a parameter mutation as well as the degree of mutation are external properties, which have a strong influence on the convergence of the algorithm. Therefore their values are adapted to the optimization task.

The correlation 
\begin{equation}
	c(X,F) = \sum_{i=0}^{N}\frac{(v\phd{i} - \braket{v})(F\phd{i}-\braket{F})}{\sqrt{(v\phd{i} - \braket{v})^2(F\phd{i}-\braket{F})^2}} 
\end{equation}
yields additional insight in the optimization dynamic and helps determining proper mutation probabilities.
Here, $F\phd{i}$ is the fitness on the basis of a set of $N \in \mathbb{N}$ measurements $X = \left\lbrace v\phd{i} | i= 1,..,N\right\rbrace$ with the mean denoted by $\braket{\cdot} $.

Specifically, the probability to mutate a parameter should be chosen according to the correlation, since for small changes it is a measure for linear response.
An absolute value $|c| = 1$ indicates a complete linear relation.
The correct choice of the mutation range is rather difficult. Ideally, the EA explores the whole parameter space. However, if in a single step the mutation range spreads over the whole parameter space, the mutation process effectively realizes a random parameter generator which does not necessarily converge on the optimization of the fitness. We privilege a small mutation range and hence we expect a linear dependence on the fitness, resulting in a vanishing correlation around local maxima. In this case, the mutation range should be extended by setting the range anti-proportional to the correlation thus allowing to leave the local maximum and spread for global optima.

After the mutation step, the fitness of the population, which comprises the newly created individuals $\menge{B}$ and their ancestors $\menge{A}$, is determined.
Experimentally, this implies a realization of the experimental cycle for every parameter set of any individual and measuring all quantities that compose the fitness, such as the atomic loading rate of a magneto-optical trap (MOT), the atom number in the dipole trap, or the phase-space density of a BEC.
Finally the EA selects those individuals with the highest fitness as ancestors for the next generation (step 4), or it stops if a termination condition such as maximum fitness is reached or total duration is exceeded (step 5).
While the bare EA randomly explores the parameter space, being rather a stochastical process, there exist many implementations that combine the power of EAs to scale with increasing dimension and approach global optima with other optimization techniques that converge faster to local maxima.

\subsection{Differential evolution (DE)}
A common way of manual optimization is to keep track of important parameters and separately tune them to an optimum. 
The DE employed in our system (see figure \ref{fig:steepest_descent}) is an adaption of a steepest descent linear optimization technique~\cite{Price2005}, capable of accelerating the convergence to a local optimum~\cite{Das2005}.
This step is added to the EA between determining the fitness of a generation and  selecting the fittest individuals.
It compares the fitness of an individual $\vec{v}\phd{i}$ to the fitness of the population in the previous generation. If the individual shows an improvement \mbox{$\Delta F\phd{i,j} = F(v\phd{i}) - F(v\phd{j}) > 0$} compared to the one of its ancestors $\vec{a}\phd{j}$ the total optimization vector 
\mbox{$\delta \vec{k} = \sum\phd{j} \Delta F\phd{i, j} (\vec{v}\phd{i} - \vec{a}\phd{j}) / \sum\phd{j} |\Delta F\phd{i,j}|$} is calculated.
Therefore, we sum over all vectors from ancestor to individual weighted with the improvement $\Delta F\phd{i,j}$.
Once such an optimization vector has been found, new individuals $\menge{O} = \lbrace \vec{o}\phd{k} = \vec{v}\phdt{i} + k \cdot \delta \vec{k} \rbrace$ following the vector are successively injected into the EA and their fitness is evaluated. In order to determine the quality of the DE vector $\delta \vec{k}$, the correlation between the DE step $k$ on the one hand and the fitness increase on the other hand is chosen. For small steps $\delta \vec{k}$ the correlation can be understood as a measure of a linear gradient, which is expected to vanish when approaching a local optimum. This allows to define a pre-selected threshold, that determines whether the EA further follows this vector or it is rejected.
The adaption of this threshold depending on the signal to noise ratio decreases the sensitivity to measurement noise, thus increasing the capability to find local optima when optimizing a noisy fitness.

\begin{figure}[h]
	\includegraphics[width=0.48\textwidth]{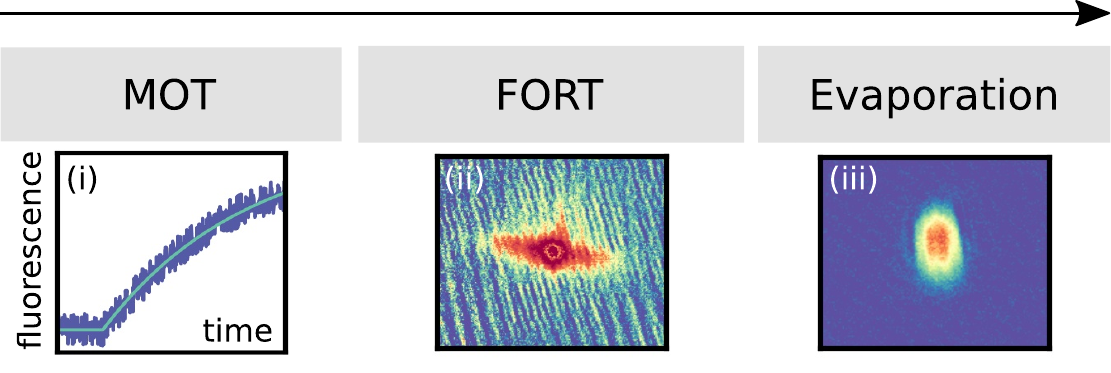}
	\caption{
		Exemplary sketches of different measurement signals that serve as fitness in the steps of the BEC creation to which we apply the EA.
		(i) Fluorescence signal from loading of a MOT. (ii) Absorption image of the atoms loaded to a crossed FORT. (iii) Absorption Image of a BEC which is optimized to a maximized phase-space density.
	}
	\label{fig:experiment}
\end{figure}

\section{Application of the EA to BEC Production}
\label{sec:application}

We apply our EA to the individual steps of our BEC creation sequence (see figure \ref{fig:experiment}) comprised of (i) laser cooling of an atomic beam in a two-dimensional (2D) MOT and subsequent trapping in a three-dimensional (3D) MOT \cite{Steane1991,Metcalf1999}, (ii) transfer of atoms to a crossed far off resonance trap (FORT) \cite{Grimm2000}, and (iii) evaporation of the sample in this optical trap~\cite{Ketterle1999}. In each step the EA has been granted full access to important parameters such as laser detuning, laser intensities, coil currents and it is able to control the timing of, e.g., laser pulses or intensity ramps. In all cases, we define a proper fitness for each individual step and extract the correlation between a parameter and the respective fitness.
\begin{figure}[b]
	\centering
	\includegraphics[scale=0.75]{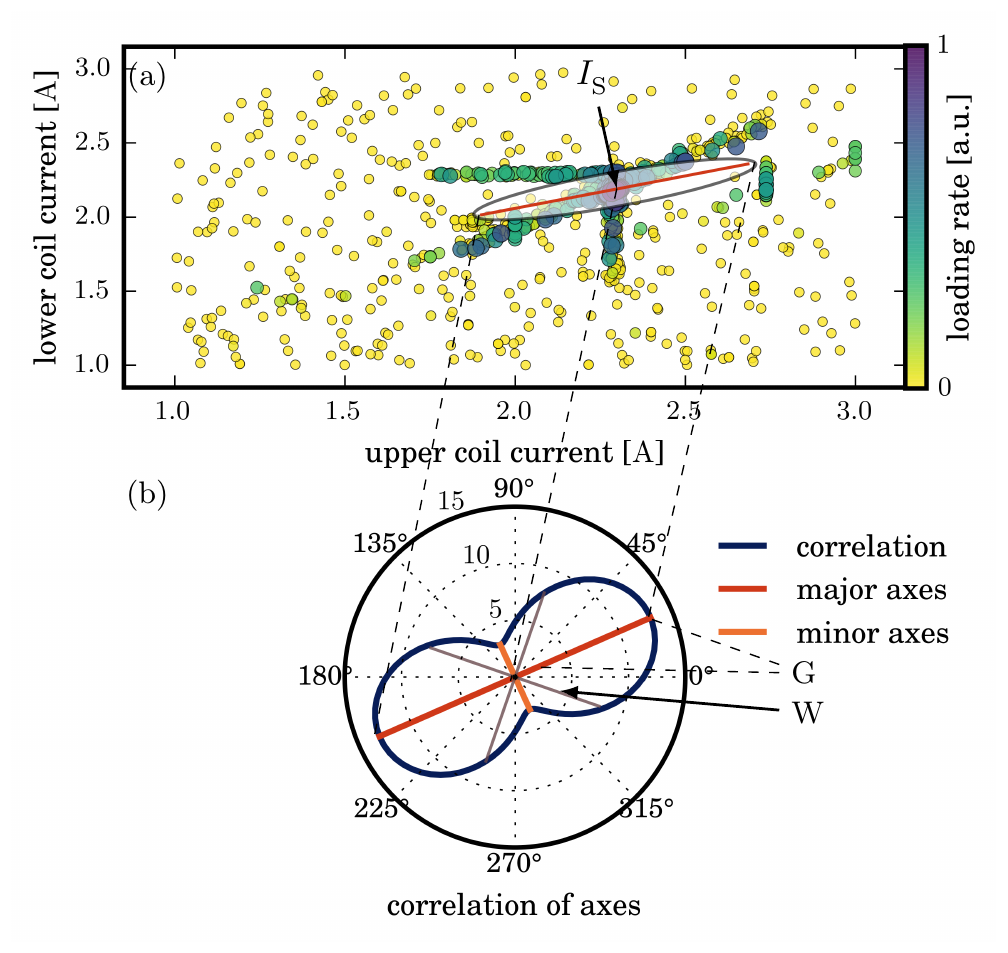}
	\caption{
		Optimization of 2D MOT performance.
		\poi{a} MOT loading rates being the fitness of optimization for vertically opposing coil currents. The fitness is encoded in color and size of the dots. Point $I\phdt{S}$ marks the center-of-mass of all data points. It is the center of the new coordinate system of the PCA, shown in \poi{b}. Tagged with \poi{G} are the main and the minor axes, which differ most in correlation. The orthogonality is due to our geometry rather than a general property of PCA's main axes. By choosing this set of axes and keeping the minor axis fixed almost no information is lost. The set of axes tagged with \poi{W} exemplifies a case where the correlation to the fitness does not differ and therefore no reduction in dimension is possible without loosing information.
	}
	\label{fig:optimization_2d_pca}
\end{figure}

\begin{figure*}[t]
	\includegraphics[width=\textwidth]{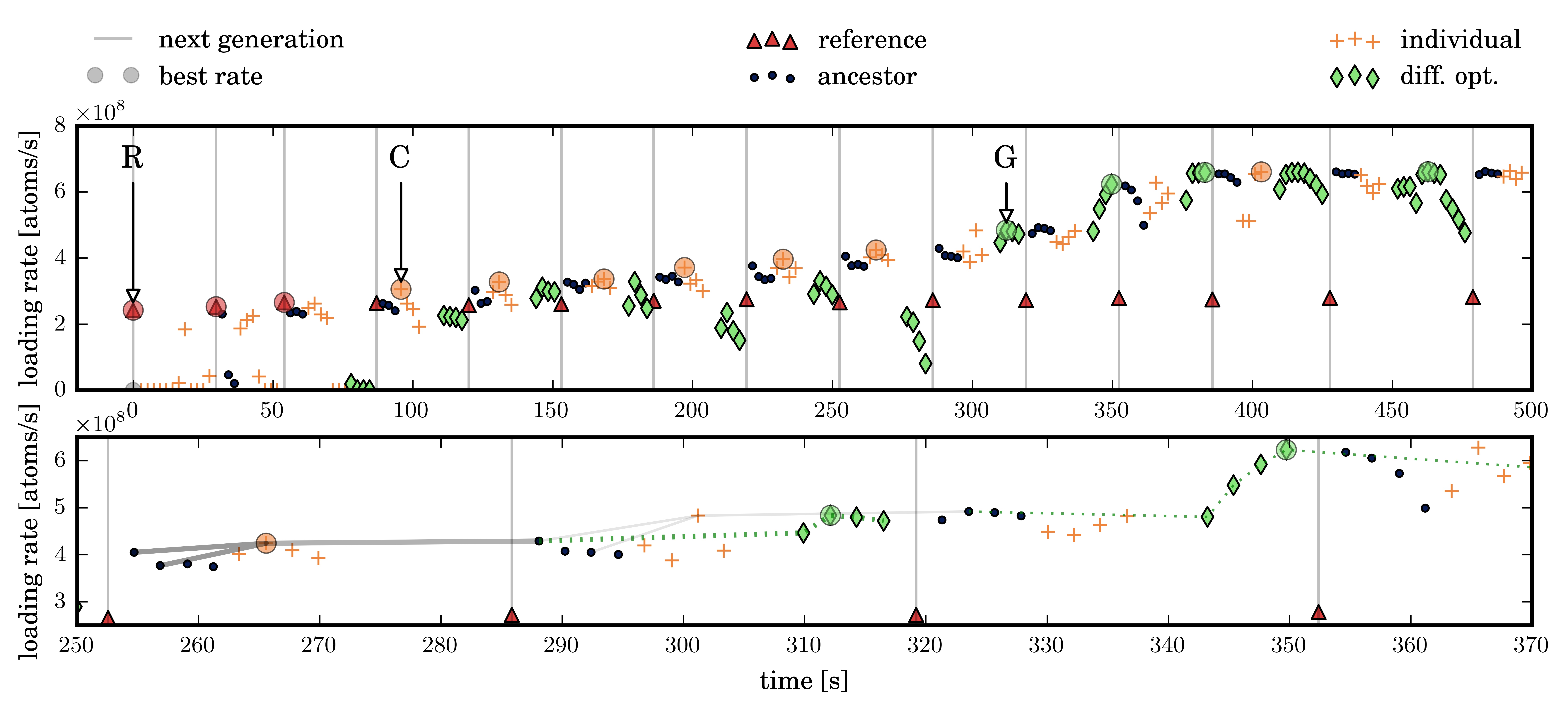}
	\caption{
		Typical time evolution of the fitness defined as MOT loading rate $\gamma$ evolving in the EA, starting from randomly chosen individuals. The reference measurements \poi{R, red triangles}, obtained from manual optimization, are not participating in EA. Generations are separated by vertical lines.
		In the first generation of the EA a single random choice reaches a significant loading rate, yet still smaller than the reference set. This individual survives the first generation and is remeasured in the following iteration, marked as survivor (blue dot).
		After $\SI{60}{\second}$ the DE is applied for the first time (green diamonds). At \poi{C} in the fourth generation the recombination resulted in individuals with a fitness exceeding the initial, manual optimization \poi{R}.
		DE further increases the loading rate from $\SI{4.0 \pm .1 e8}{\atoms \per \second}$ to $\SI{6.6 \pm .1 e8}{\atoms \per \second}$ at \poi{G}. The genealogy of the differential optimization \poi{G}, which is plotted below, increases the power of the cooling laser and correlates up to $c = 0.92 $ with the fitness. Details of this optimization vector are given in table \ref{tb:optimizsation_3d_from_scratch}.
	}
	\label{fig:optimization_3d_from_scratch}
\end{figure*}
\subsection{(i) MOT optimization}
In a first step of the BEC creation sequence a large number of atoms are laser cooled and captured in a 3D MOT. The 3D MOT is loaded from a pre-cooled atomic beam originating from a 2D MOT, described in \cite{Hohmann2015}.
To optimize the 3D MOT loading, where the atom number of trapped atoms $N$ can be described by an exponential $N(t) = N_\mathrm{max} \cdot (1- \exp\{- \gamma t \})$, we obtain the loading rate $\gamma$ from a fit.
The parameters to be optimized thus include the laser detunings and intensities for the 2D and 3D cooling and repumper lasers, the intensity and detuning of a push beam which guides the atoms from the 2D to the 3D MOT region, and the currents of the magnetic quadrupole fields of the MOT. In total the dimension of this optimization problem sums up to 12 parameters.
Specifically, the quadrupole fields for the 2D MOT are produced by four coils, the currents of which can be independently adjusted. Here, the ratio of opposing coil currents determines the position of the 2D MOT, while the absolute current determines the size of the cooling area. The atomic beam emitted from the 2D MOT has to pass a differential pumping tube with a diameter of $\SI{1.8}{\milli\meter}$. Thus, the ratio of opposing currents is rather fixed by this constraint, effectively allowing the parameter space to be reduced.
We use a mathematical tool called principal component analysis (PCA) \cite{Handl2010} in order to identify a rotated coordinate system in parameter space which is more suited for optimization. Specifically, we aim for finding major axes of the rotated coordinate system that contain important information of the problem, while minor axes can be neglected with minimal loss of information. The amount of information of an axis is described by the correlation of the fitness to variations along the axis.
We apply a PCA to the fitness values obtained from the optimization of our 2D MOT coil currents.
The result is depicted in figure \ref{fig:optimization_2d_pca} for opposing coils and clearly shows a high correlation on a line close to a current ratio of unity, with a starting point $I\phdt{S}$ for further applications of the EA. The perpendicular axis, in contrast, only shows a weak correlation. This is an optimal coordinate system for our problem with a major axis along the high correlation, and we neglect the weakly correlated minor axis in the following. This coordinate system especially improves the reproduction process since we use a discrete inheritance which does not necessarily conserve the ratio of parameters, e.g., opposing coil currents.

Now we add the parameters of the 3D MOT to the EAs optimization task and  improve the loading rate further.
An increase in fitness value will show the progress of the EA, however, the fitness can also be affected by drifts of the experimental apparatus.
In order to quantify the progress of the EA independently from these drifts, we apply reference measurements with a predefined parameter set, representing the best manual optimization. This parameter set is evaluated at the beginning of each new generation but it is not taking part in any population of the EA. Figure \ref{fig:optimization_3d_from_scratch} compares the evolution of the fitness for the 3D MOT with such reference measurements. The EA starts from random initial parameters with initially vanishing fitness. The graph demonstrates the typical observation, where the EA exceeds the fitness of the manually obtained optimum by a factor of three after $\SI{400}{\second}$. Importantly, both components of our algorithm, evolutionary strategy as well as differential optimization, contribute to the total improvement.

The parameter values of the differential optimization depicted in figure \ref{fig:optimization_3d_from_scratch} are given in table \ref{tb:optimizsation_3d_from_scratch}. 
Here, we observe an increase in loading rate that is highly correlated ($c = 0.93$) to the increase of the cooling laser power. 
However the expected parameter, \ie the cooling laser intensity control value, surprisingly remains unchanged.
This indicates that the EA has found a hidden correlation. Pathologically, a frequency change of our laser lock setup also affects the intensity of the diffracted laser beam in the acousto-optical modulators (AOM) that are used for fast switching of intensities and detuning the laser frequencies. This is caused by modifying the working point of the AOM on the non-linear curve of diffraction efficiency versus frequency. Thus, the AOM output power depends on the frequency chosen. 

\subsection{(ii) Atom transfer to the FORT}
From the MOT, the atoms are transfered to a crossed FORT at $\SI{1064}{\nano\meter}$ for evaporative cooling to degeneracy.
To optimize the initial conditions for the evaporation, the transfer rate is improved with our EA, choosing the total amount of transferred atoms as fitness.
Here, we add an initial phase of self evaporation where the laser intensities are kept constant and an additional compressed-MOT (CMOT) phase \cite{Ketterle1993,Lewandowski2003}, where the cooling laser is far red detuned leading to a decreased heating from re-emission of photons.
This results in a anti-correlation of $c = -0.09$ between the cooling laser power and the fitness, which means that an increase in cooling laser power results in a lower loading rate.
In contrast to the MOT, the repumping laser intensity here is highly anti-correlated with $c \approx -0.74$, since less repumping power allows more atoms to leave the cooling cycle, which minimizes heating.
This optimization increases the total atom number loaded to the FORT from $\SI{2.6e5}{\atoms}$ to $\SI{1.05 e6}{\atoms}$ by a factor of four and is used as starting point for the following evaporation process.

\subsection{(iii) Optimization of the evaporation}
As a last step we focus on the experimental details and analyze the consequences when applying the EA to optimize the BEC production, namely evaporation \cite{Lewandowski2003,Clement2009}, and improve the phase space density.
The optimization parameters for the evaporation process include initial and final intensities as well as the decay times of the exponentially ramped trapping potentials for both arms of the crossed FORT. Furthermore we introduce a delay between the start of both evaporation ramps. 
In order to estimate the phase space density we image the atomic sample with a time-of-flight (TOF) of \mbox{$\tof = \SI{16}{\milli\second}$}. We extract the widths $\sigma_{x,y}$ and the total atom number
$N = A \left[\pi (\sigma_x^2+\sigma_y^2)\right]^{1/2}$, where $A$ is the optical peak density integrated over the $z$-direction. For the fit of the measured distribution we use a Gaussian distribution rather than a bimodal fit, which is numerically unstable when spanning a large parameter space across the critical point of Bose-Einstein condensation.
Thereby we estimate the temperature $T \propto (\sigma_x^2 + \sigma_y^2) / \tof$, neglecting the clouds in trap size, justified by the large TOF.
The fitness for the algorithm is defined as the peak density $\sfrac{A}{(\sigma_x^2 + \sigma_y^2)}$, which is a coarse approximation of the phase space density $\rho = n \lambda^{3}_\text{dB}$, where $\lambda\phdt{dB} \propto T^{\sfrac{-1}{2}}$ is the thermal De-Broglie wavelength and $n = \sfrac{N}{V\phdt{Trap}}$ the atomic density.

\begin{figure}[b]
	\includegraphics[width=.48\textwidth]{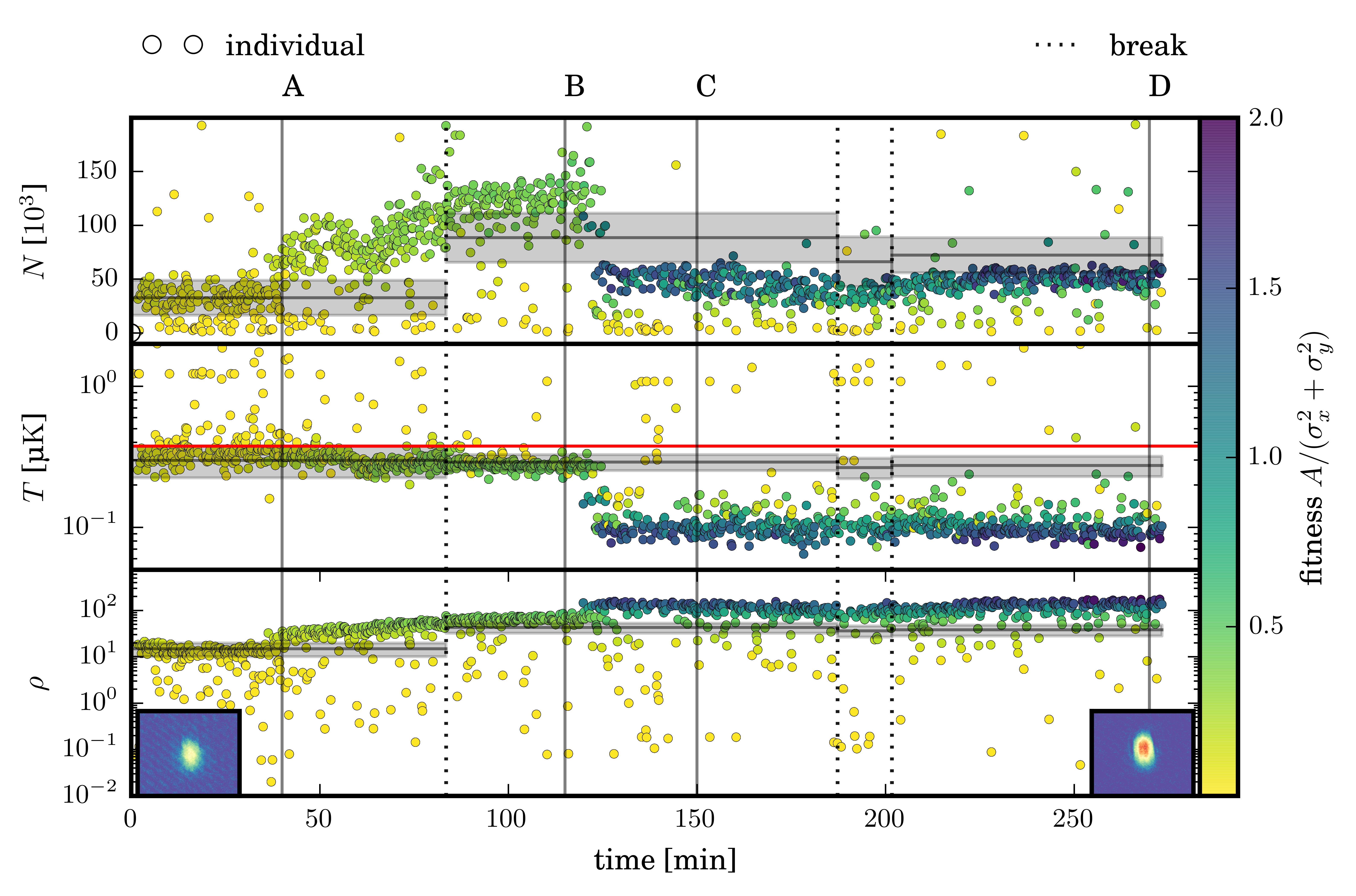}
	\caption{
		Optimization of evaporative cooling to a maximum phase space density $\rho$. The color code visualizes the obtained fitness which is directly proportional to the phase space density. The mean values of the reference measurements are marked by a black line with its confidence interval (shaded area). Since the reference parameter set has been readjusted to new optimum values after each break (dashed vertical line) it is not continuous. The different regions separated by vertical solid grey lines indicate, where the EA follows different optimization strategies \region{A-D}.
		The dotted vertical lines indicate a break in optimization, the red horizontal line is the critical temperature.
		The insets show cold clouds at the beginning (left inset) and at the end (right inset) of the optimization process, where color coding is the same.
	}
	\label{fig:optimization_bec}
\end{figure}

The evolution of the evaporation is depicted in figure \ref{fig:optimization_bec}, where additionally to the fitness the absolute atom number and the temperature are shown.
In the region \region{A-B} of figure \ref{fig:optimization_bec}, the algorithm increases the atom number $N$ while in \region{B-C} it reduces the temperature $T$, both leading to an increased phase space density $\rho$. 

By correlating applied changes to measured observables, we determine the parameters with the strongest impact.
In the region $\region{A-B}$ the increase of atom number has a maximum correlation of $c = 0.46$ to the CMOT duration.
Extending the CMOT duration and consequently decreasing the duration of the self evaporation in the FORT increases the number of atoms loaded results in an augmentation of remaining atoms after the evaporation.
On the one hand, the phase space density is directly proportional to the atom number, hence increasing the CMOT duration is a good optimization vector. On the other hand, we have limited the time for the CMOT duration to avoid violation of timing constraints.  At \poi{B} in figure \ref{fig:optimization_bec} the optimization vector violates timing constraints and therefore cannot increase the atom number any further. This results in a positive correlation indicating that further improvement by extending the CMOT duration is possible.

In region $\region{B-C}$ we observe that the atomic temperature is decreased from \poi{B} $T=\SI{0.29 \pm 0.05}{\micro\kelvin}$ to \poi{C} $T=\SI{0.10 \pm 0.05}{\micro\kelvin}$.
We find that the evaporation ramp decay time has the highest absolute correlation $c \approx -0.48$ which is therefore the main reason for the temperature drop.
Decreasing the decay time results in less time for the atoms to thermalize and  increased loss of atoms during the evaporation process, causing the average atom number to drop from \poi{B} $N = 101(28)\cdot10^3$ atoms to \poi{C} $N = 46(12)\cdot 10^3 $ atoms.

Besides yielding information about important parameters for the evaporation and FORT loading processes, the application of the EA resulted in an increased phase space density which enabled us to reduce the time for the creation of our BEC from $\SI{8}{\second}$ to $\SI{4}{\second}$.

\section{Conclusion}
\label{sec:outlook}

We have discussed the implementation of an EA enhanced with a differential evolution method to optimize the different steps of BEC production. Particularly, we introduced the various fitness definitions we applied to improve these individual trapping steps. We showed that the parameter space can be reduced by applying a principal component analysis and furthermore illustrated the genealogy of the EA, that reveals non trivial correlations of the experimental setup.
Basically the EA can be applied to any optimization problem and in our application resulted in several improvements such as the increase of magneto-optical trap loading rate by a factor of $\approx 3$, an increase in total atom number loaded into the dipole trap by a factor of $\approx 4$, when starting with already optimized parameters from the MOT and an increase of the BEC phase space density by an order of magnitude allowing for a BEC creation in $\SI{4}{\second}$.
We have shown that, by applying the EA, we do not only gain long term statistics that yield information about drifts of devices and optimal parameter sets, but we also learn about further optimization options by analyzing correlations between parameters and fitness and by evaluating the DE optimization vectors.

\subsection{Acknowledgement}
	The project was financially supported partially by the European Union via the ERC Starting Grant 278208 and partially by the DFG via SFB/TR49.
	D.M. is a recipient of a DFG-fellowship through the Excellence Initiative by the Graduate School Materials Science in Mainz (GSC 266),
	F.S. acknowledges funding by Studienstiftung des deutschen Volkes, and
	T.L. acknowledges funding from Carl-Zeiss Stiftung.

	\bibliography{manuscript.bib}
	
\renewcommand{\thesection}{A }
\section{Supplementary}

\begin{table}[H]
	\caption[Optimization vector \poi{G} depicted in figure \ref{fig:optimization_3d_from_scratch}]{Data of the optimization vector \poi{G} depicted in figure \ref{fig:optimization_3d_from_scratch} with relevant changes (degree of mutation $\nu > 1\%$) and correlations of observables ( $c > 0.7$).
		We observe highest correlation to the 3D cooling laser power ($\coolerPower$), which is the main reason for the increase of the loading rate.
		Pathologically the intensity control voltage $\coolerIntensity$ is not changed, thus the reason of the increase in $\coolerPower$ is a consequence of a non trivial hidden correlation of our setup.
		The spectroscopy $\spectroscopy$ is the first of two AOMs, that control cooling lasers detuning.
		In order to keep the 3D MOT detuning $\coolerDetuning$ constant the subsequent AOM compensates the change per linear optimization step of $\delta\vec{k}(\spectroscopy) = -\MHzPi{3.0}$ and thereby optimizes the working point resulting in an increased $\coolerPower$.
	}
	\label{tb:optimizsation_3d_from_scratch}
	\centering
	\begin{tabular}{ l|c c c }
		\toprule
		parameter		 	& start 		&  $\delta\vec{k}$		& \\
		\midrule
		\spectroscopy 		& $\MHzPi{435.7}$ 	& $-\MHzPi{3.0} $	& \\ 
		\coolerdetuning 	& $-\MHzPi{13.5}$ 	& $+\MHzPi{0.3} $	& \\
		\toprule
		observable 			& start 		& change                 & corr. \\
		\midrule
		$\coolerPower$ & $\SI{118}{\milli\watt}$ & $+\SI{23}{\milli\watt}$ & $\num{0.93} $ \\
		$\coolerpower$ &$\SI{79}{\milli\watt}$ & $+\SI{3.39}{\milli\watt}$ & $\num{0.83} $ \\
		$\repumpPower$ & $\SI{2.58}{\milli\watt}$ & $+\SI{48}{\micro\watt}$ & $\num{0.72} $ \\
	\end{tabular}
\end{table}

\end{document}